# Decrystallization of $CH_3NH_3PbI_3$ perovskite crystals via polarity dependent localized charges

*Min-cheol Kim[a†], Namyoung Ahn[a†], Eunhak Lim[b], Young Un Jin[c], Peter V. Pikhitsa[a], Jiyoung Heo[d], Seong Keun Kim[b], Hyun Suk Jung[c*] and Mansoo Choi[a,e*]*

[a]Global Frontier Center for Mulitscale Energy Systems, Seoul National University, Seoul, Republic of Korea.

[b]Department of Chemistry, Seoul National University, Seoul, Republic of Korea.

[c]School of Advanced Materials Science & Engineering, Sungkyunkwan University, Suwon, Gyeonggi-do, Republic of Korea.

[d]Department of Green Chemical Engineering, Sangmyung University, Seoul, Republic of Korea.

[e]Department of Mechanical Engineering, Seoul National University, Seoul, Republic of Korea.

[†]These authors contributed equally to this work.

[*]Correspondence and request for materials should be addressed to H.S.J, and M. C.

(email: hsjung1@skku.edu; mchoi@snu.ac.kr)

# Abstract

Despite soaring performance of organic-inorganic hybrid perovskite materials in recent years, the mechanism of their decomposition at actual operation condition has been elusive. Herein, we elucidated the decrystallization process of $CH_3NH_3PbI_3$ perovskite crystals via localized charges and identified polarity-dependent degradation pathway by carrying out time-evolution measurements for absorption spectra of perovskite films with underlying different charge transport layers and *ab initio* molecular dynamics calculations. It was found that the carrier polarity (hole-rich or electron-rich) inside the perovskite films played a critical role in the degradation rate, and polarity-dependent degradation pathway strongly depended on the combination of surrounding gaseous molecules. The hole-rich perovskite films degraded more rapidly in the existence of $H_2O$ than the electron-rich one, while the degradation trend became opposite in only-oxygen ambient. Strikingly, the hole-rich one was extremely weak to atmospheric air containing both $H_2O$ and $O_2$, whereas the $MAPbI_3$ film with excessive electrons rather stabilized in air. *Ab initio* molecular dynamics (AIMD) simulation was also done to find the detailed degradation pathway of $MAPbI_3$ under atmosphere for different polarity of localized charge, which are in good agreement with experimental results. Furthermore, X-ray assisted spectroscopic measurements confirmed the production of $Pb(OH)I$ as predicted from the simulation result.

Perovskite is an ionic crystal with three kinds of ions, in which the cation and anion are being held together by electrostatic forces. As one of perovskite materials, organic-inorganic hybrid perovskites have currently attracted worldwide attention due to their excellent performance for photovoltaics,[1-4] but those have shown unstable characteristics against light and air.[5, 6] Since such hybrid perovskites have weaker hydrogen bonding between organic cation and octahedral structure than inorganic ones,[7] organic cations have been considered to be responsible for easy crystal de-bonding and chemical reaction with atmosphere.[8, 9] Hybrid perovskites rapidly change into inorganic halide after light-induced degradation.[10] Such fast decomposition of hybrid perovskites impedes to commercialize perovskite-based photovoltaics.

Degradation of these perovskite materials have been studied for past few years,[11-14] however, there remain a lot of unknowns and debates on degradation mechanism because the process happens under extremely complicated multi-molecular systems in real operation condition.[10, 15, 16] As possible causes, researchers previously suggested that ion migration of halide anion and resulting deprotonation of organic cation.[5, 11, 17, 18] However, it is possible that observed ion migration and deprotonation could be merely a result of degradation, not the origin of degradation. Accordingly, digging out the fundamental reason of degradation is highly urgent to solve the stability issue.

Our group had pointed out trapped charges in perovskite ionic crystals as the fundamental origin.[19, 20] We experimentally demonstrated that trapped-charges along grain boundaries and also on grain surfaces induced irreversible degradation in the presence of water, oxygen or air. We also carried out *ab initio* molecular dynamics (AIMD) simulation for $CH_3NH_3PbI_3$ (MAPbI$_3$) crystals with adsorbates of either H$_2$O and O$_2$ to examine whether the trapped charges could trigger the degradation of the MAPbI$_3$. From these experiments and simulations, we uncovered atomistic mechanism for trapped-charge driven degradation of

MAPbI$_3$, in which trapped-charges play a decisive role in breaking MAPbI$_3$ lattices as a consequence of intermolecular interaction with gaseous molecules under light soaking. However, it still remains unknown whether the polarity of trapped charges could result in different pathways of the degradation of perovskite materials under the interaction with multi-molecular species in the real operation condition.

In the present work, we found a strong evidence to support our trapped-charge driven degradation and identified its polarity dependent decrystallization process through light-induced degradation test of MAPbI$_3$ perovskite films coated on different charge-selective layers.[21, 22] The hole- and electron-rich MAPbI$_3$ layer can be realized in half devices which were fabricated by coating selective charge transporting underlying layers. The existence of unbalanced hole and electron densities was evidenced by photoluminescence (PL), Kelvin force microscopy (KPFM) measurements and theoretical calculations. These electron or hole-rich half devices were exposed to different gaseous molecules to dig into the role of polarity dependent localized charges on the degradation. In moisture-only ambient, the degradation rate of the hole-rich MAPbI$_3$ film was faster than that of electron-rich one, while oxygen-only case presented totally-inverted results with respect to the degradation rates. More importantly, the actual atmospheric conditions (when oxygen and moisture exist together) were highly aggressive only for the hole-rich MAPbI$_3$ film, whereas the electron-rich one rather stabilized (even more stable than oxygen-only case). Such interesting results suggest that there exists a direct correlation between polarity dependent localized charges and degradation. To explain why and how charge polarity influences the degradation rates depending on environments, we conducted AIMD simulations to study intermolecular interactions between MAPbI$_3$ components and mixture gas (oxygen and water) with different charge state (charge polarity of MAPbI$_3$). In the simulations, under the mixture of oxygen and water, it was found in the hole-rich case (with positive charge) that strong interaction of Pb-O bonding induces the vigorous destruction of $PbI_6$ octahedron and finally generates hydroxide species are formed as a consequence of proton transfer from $H_2O$, which was experimentally confirmed from X-ray assisted spectroscopic measurements in the present study.

**Results**

**Different degradation rates of MAPbI$_3$ depending on underlying transport layers**

We employed perovskite films with different underlying layers to investigate the effect of underlying layers on degradation rates under light illumination in the same ambient condition. We configured the half device with ITO/selective charge extraction layer/MAPbI$_3$, in which surfaces of MAPbI$_3$ films were directly exposed to air without any passivation layer, and this configuration enables films to be exposed to the same ambient condition under light illumination. As we will discuss in detail later, half devices have unbalanced polarity of charge carriers inside the MAPbI$_3$ films (due to selective charge extraction and consequently cause unbalanced surface localized charges confirmed by SKPM.[23]) We carried out degradation tests with different ambient conditions by using our customized experimental setup that can keep specific ambient condition inside the chamber. (**Fig. 1a**) Interestingly, the degradation of the PEDOT:PSS/MAPbI$_3$ half device which has excessive electrons happened much faster than that of the C$_{60}$/MAPbI$_3$ half device which has excessive holes in dry air ambient (N$_2$(80%)+O$_2$(20%)) (**Fig. 1b, c**), while humidified nitrogen ambient (N$_2$+H$_2$O R.H.~80%) caused totally inverted results, i.e. in this case, C$_{60}$/MAPbI$_3$ half device containing excessive holes degraded much faster than the case of PEDOT:PSS/MAPbI$_3$ (**Fig. 1d, e**). Under 100% oxygen ambient condition, the degradation occurred more rapidly than dry air ambient, but exhibited the same trend depending on the underlying layers (**Supplementary Fig. 1**). It is noteworthy that the PEDOT:PSS/MAPbI$_3$ half device shows very slow degradation in the humidified nitrogen ambient in contrast to the C$_{60}$/MAPbI$_3$ half device. When surrounded only by nitrogen, no light-induced degradation of the MAPbI$_3$ film was observed regardless of the underlying transport layers (**Supplementary Fig. 2**), which is consistent with previous studies.[5, 19] This means that the underlying transport layers themselves cannot deteriorate the

MAPbI$_3$ films even with unbalanced localized charges.

Previously, it was found that MAPbI$_3$ films were decomposed under light illumination in dry air ambient even without moisture.[5, 11] This study demonstrated that superoxide (O$_2^-$) formed by the reaction of oxygen molecules with photo-generated electron could degrade MAPbI$_3$ perovskites without water molecules. The authors claimed that, in the case of the TiO$_2$/MAPbI$_3$ half device which can selectively quench electrons, superoxide generation decreased as compared to the case of either the glass/MAPbI$_3$ or Al$_2$O$_3$/MAPbI$_3$ half device, which finally led to relatively slow degradation. Our experimental results also present similar results in dry air ambient (**Fig. 1b, c**). However, the degradation rate in humidified nitrogen ambient was completely reversed, which has not been known before. These observations clearly show strong dependency of degradation rate on the kind of underlying layers and surrounding ambient conditions. Obviously, different charge transport layer changes the distribution of charge carrier polarity inside the MAPbI$_3$ layer.

**Unbalanced charge carrier polarity inside MAPbI$_3$ layers**

To explain such interesting phenomena, it is required to analyze how the kind of underlying layers leads to difference in degradation rates. The kind of underlying layers influence carrier density distributions in the MAPbI$_3$ layer under light illumination.[23] We prepared four different half devices to examine the change in carrier densities depending on the underlying layers via photoluminescence (PL) measurements. We used C$_{60}$[24] and PC$_{60}$BM[25] as ETLs, and PEDOT:PSS[26] and Spiro-MeOTAD[27] as HTLs for half device configurations. Since radiative recombination is proportional to electron and hole densities,[28] PL intensities are indicative of carrier densities in MAPbI$_3$ layers. Dramatic decreases in PL intensities were observed with the case of either electron transport layer (ETL)/MAPbI$_3$ or hole transport layer (HTL)/MAPbI$_3$ half devices compared to the case without transport layers, which indicates that photo-generated electrons and holes are selectively collected at the interface, respectively (**Supplementary Fig. 3**). In the Spiro-MeOTAD case, additives such as TBP and Li-TFSI were not added to avoid of controversial issue on the effect of their corrosive properties.[29, 30] Its charge extraction property was not significantly influenced by the absence of additives, as indicated by the results of PL measurements (**Supplementary Fig. 4**). In the same manner, time-resolved PL data also presented strong carrier extraction properties of ETL/MAPbI$_3$ and HTL/MAPbI$_3$ half devices (**Supplementary Fig. 3**). The values of $\tau_1$ (the fastest decay component) for these half devices, which correspond to charge injection,[31] were far smaller than that for the glass/MAPbI$_3$ film as can be seen in Supplementary Table 1. In other words, these results imply that such charge-selective layers strongly attract a particular carrier from MAPbI$_3$ films at the interfaces.

Based on PL results, we can assume that electron and hole density at the interface is zero in the case of ETL/MAPbI$_3$ and HTL/MAPbI$_3$ device, respectively. By solving carrier continuity equation considering light absorption, carrier mobility, carrier recombination, carrier drift and diffusion, we calculated electron and hole densities as a function of thickness displacement under one sun illumination for different underlying layers. (**Calculation details in Supplementary Information**). **Fig. 2c-e** show calculation results of electron ($n_e$, indicated in red line) and hole ($n_h$, indicated in blue line) density distributions depending on different underlying layers. The glass/MAPbI$_3$ device have uniform distributions due to the absence of the charge selection layer, in which the orders of magnitude in hole and electron density are similar. Small difference originates from mobility difference between electron and hole.[32] On the other hand, in the ETL/MAPbI$_3$ device, hole densities are much higher than electron densities due to electron quench at the ETL interface. Strong and fast electron transfer from MAPbI$_3$ to ETL is inferred from a steep slope of electron densities near the ETL interface.

Instead, the HTL/MAPbI$_3$ device have inverted carrier distributions. It is also confirmed not only that hole densities dramatically change near the HTL interface like the ETL/MAPbI$_3$ case, but also that electron density is higher than hole density. It is clearly shown that the kind of underlying layer results in the unbalance of carrier polarity inside MAPbI$_3$ film when photo-carriers are generated by light illumination. Additionally, we performed Kelvin probe force microscopy (KPFM) measurements to experimentally demonstrate unbalance of charge carrier.[33] Imbalance carrier concentration inside the MAPbI$_3$ layer can alter quasi fermi level as depicted in **Supplementary Fig. 5**, which correspondingly could change surface potential. **Fig. 2f-h** clearly show different surface potentials depending on the underlying layer, which is in good agreement with density distribution calculations. Since hole densities are much higher than electron densities in the ETL/MAPbI$_3$ half device regardless of thickness displacement, the device shows the highest potentials among three kinds of half devices (**Fig. 2i**). Based on the trapped charge driven degradation mechanism[19], we can hypothesize that the different degradation phenomena for different charge selective layers would be attributed to different charge polarity (electron-rich or hole-rich) inside the MAPbI$_3$ films.

**Correlation between degradation rates and charge polarity under different ambient conditions**

Since we investigated the light-induced degradation of half device configurations under the oxygen-only or humidified nitrogen-only condition so far, we additionally performed the degradation experiments under atmospheric air condition containing both oxygen and water. (see **Fig. 3a, b** and **Supplementary Fig. 6.**) We observed unexpected outcomes from these experiments in terms of degradation rates. The complete deterioration of the ETL/MAPbI$_3$ device took place within 12 hours as shown in **Fig. 3a**. The degradation rates were by far faster than those under either dry air or humidified nitrogen ambient (see **Fig. 3c**). Namely, the mixture of oxygen and water violently destroyed the ETL/MAPbI$_3$ devices, as compared to the case of only-oxygen or only-water ambient. On the other hand, the HTL/MAPbI$_3$ devices showed surprisingly enhanced stability under humidified air as seen in **Fig. 3b**. The remarkable point of observations in case of HTL/MAPbI$_3$ is the fact that the degradation under air with moisture occurred more slowly than the degradation either under air without moisture or humidified nitrogen without oxygen (**Fig. 3d**). Surprisingly, an addition of moisture rather improved the stability of the HTL/MAPbI$_3$ device. Consequently, the mixture of oxygen and water gives rise to extremely diametrical effects on degradation rate depending on the kind of underlying layers (different excessive carrier polarity). The origin of such different degradation rate is further studied in atomic scale by AIMD simulation discussed below.

The X-ray diffraction (XRD) patterns obtained for the pristine and illuminated MAPbI$_3$ films deposited on different charge transport layers supports the observation (**Supplementary Fig. 7**). The MAPbI$_3$ films with localized hole were characterized by the stronger intensities of their PbI$_2$ (100) peaks as compared to those with localized electron samples, which was in good agreement with the results of absorption studies. The observed change in the PbI$_2$ (100)/MAPbI$_3$ (110) peak intensity ratio before and after illumination is summarized in **Supplementary Table 2**. The obtained scanning electron microscopy (SEM) images demonstrate the same light stability trends that were observed during the absorption and XRD studies (**Supplementary Fig. 8** and **Supplementary Fig. 9**). These results from optical, spectroscopic and morphological analysis indicate that gaseous molecules and their combination result in different degradation pathways depending on localized charge polarity of MAPbI$_3$ crystals.

We also investigated balanced charge carrier density case by testing MAPbI$_3$ film with no transporting layer (glass/MAPbI$_3$). In the case, the light-induced degradation happened at a similar rate to the ETL/MAPbI$_3$ as shown in **Supplementary Fig. 6**. Even if the MAPbI$_3$ have balanced carrier densities of electrons and holes, the magnitude of carrier densities is much larger than those of the ETL or HTL/MAPbI$_3$ (**Fig. 2c-e**), which induced faster degradation than expected. Moreover, since MAPbI$_3$ crystals might have locally unbalanced carrier densities at defect sites such as grain boundaries,[34] high carrier densities of holes can cause such fast degradation of the glass/MAPbI$_3$ film under air conditions.

**Discussion**

**Different degradation rates of MAPbI$_3$ under H$_2$O-only and O$_2$-only ambients**

It is noteworthy that degradation rates of MAPbI$_3$ highly depend on charge carrier density and surrounding ambient. We have demonstrated trapped-charge driven degradation mechanism evidenced by experimental results and AIMD simulations in our previous works. As additional evidence for trapped-charge driven degradation, we newly discovered that unbalance of charge carrier (charge polarity) plays a different role in degradation rates in the present work. Based on AIMD simulation results in our previous work, we could find out how the charge polarity determine different degradation rates under H$_2$O-only and O$_2$-only ambient. According to simulation results, the H$_2$O-only ambient shows different solvation characteristics depending on the polarity of MAPbI$_3$ crystals.[20] Water molecules solvate methylammonium cation (MA$^+$) at the surface of MAPbI$_3$ crystal with localized hole (positive polarity), while, in the case of negative polarity, water molecules surround and solvate iodide anion (I$^-$). Since hydrogen bonding between MA$^+$ and PbI$_6^-$ is weaker than the covalent bonding of Pb and I in PbI$_6$ octahedron, MA$^+$ could be more easily solvated by water due to such weak bond strength. As a result, MAPbI$_3$ crystals with localized hole became unstable in only-moisture case due to easy solvation of MA$^+$. Relatively heavy I$^-$ anion with strong bonding is more durable against water solvation in the case of MAPbI$_3$ with negative polarity. These observations are in good agreement with our present experimental results regarding only-moisture degradation test (**Fig. 1d, e**). Next, in the only-oxygen case, MAPbI$_3$ crystals with negative polarity were broken down faster than ones with localized hole, which was attributed to highly reactive superoxide generation. In our previous AIMD simulation results,[20] generated-superoxide formed Pb-O bonding, thereby broke down PbI$_6^-$ octahedral. The different quantity of generated superoxide depending on charge polarity was revealed from these simulation results. The MAPbI$_3$ crystals with negative polarity generate more superoxides than the positive case, which is consistent with the reference reported by Haque's group.[5, 11] The difference of degradation rates from our experimental results under only-oxygen ambient (**Fig. 1b, c**) can be clearly explained by quantity difference of superoxide generation.

**AIMD simulations for MAPbI$_3$ decrystallization under the mixture of O$_2$ and H$_2$O**

We observed totally different degradation behavior in the presence of both oxygen and water (**Fig. 3a, b**), of which the reason is still elusive. Different synergetic effects of the combination of oxygen and water on degradation rate definitely indicate that there exist distinct and complicated intermolecular actions depending on charge polarity of MAPbI$_3$ crystals. Therefore, it is of high importance to investigate the mechanism of the degradation of MAPbI$_3$ at real operation condition containing both oxygen and water. In order to explore intermolecular dynamics of oxygen and water depending on the polarity of localized charges, AIMD simulations with configuration containing both oxygen and water molecules were performed for the first time. We set an initial geometry with 2 rigidly fixed MAPbI$_3$ units in the bottom, 2 relaxed MAPbI$_3$ units at the top to form the surface, and embed oxygen and water molecules into gaps between atoms. The surface of MAPbI$_3$ crystals are terminated with MA cations and I anions and given different net charges (+1, 0, or -1) in the unit cell to model localized charges. **Fig. 4a-c** show molecular dynamics of MAPbI$_3$ crystals with surrounding one oxygen molecule and 3 water molecules depending on different charge state (+1, 0, or -1). Interestingly, in the case of +1 charge, the PbI$_6^-$ octahedron at the surface is easily destroyed by interactions with oxygen molecules, while other charge states present stable octahedral structure of PbI$_6^-$ with the same surrounding molecules (See **Fig. 4a**). Destroy of PbI$_6$ octahedral structure in +1 charge state is likely to be mainly attributed to the strong Pb-O bond, which was not shown in neutral and -1 charge state. Furthermore, after Pb-O bond generation, a domino effect of proton transfer from surrounding water molecules finally results in deprotonation of MA$^+$ cation which turns into methylamine gases (750 fs of **Fig. 4a**). Such molecular dynamics in +1 charge state will be the reason why the fastest degradation of the ETL/MAPbI$_3$ device (MAPbI$_3$ with localized hole) happened in humidified air. In the neutral charge state, oxygen molecules gradually drift apart from the MAPbI$_3$ surface and water molecules do not react with any MAPbI$_3$ components (see **Fig. 4b**). Also, the -1 charge state leads to unexpected molecular interactions in that an oxygen molecule attach to iodide anion, but there is no further destroy of PbI$_6^-$ octahedron (**Fig. 4c**), which indicates that oxygen cannot directly attack Pb atom at the center of PbI$_6$ octahedron in the MAPbI$_3$ crystals with localized electron. To clearly compare intermolecular interactions in these systems, we analyzed Pb-O, I-O and O-H distances for atoms of interest during the simulation, respectively. (**Fig. 4d-f**). In the +1 charge state, Pb and I atoms of MAPbI$_3$ crystal have the strongest interactions with oxygen molecules, which is evidenced by the shortest Pb-O and I-O distance in **Fig. 4d, e**. On the other hand, there seems to be weak intermolecular interaction between MAPbI$_3$ components and gaseous molecules in both cases of neutral and −1 charge state (**Fig. 4d-f**). Also, we compared O-H distance to verify the deprotonation of MA$^+$ cation. The hydrogen and oxygen atoms are the component of MA$^+$ cation and water molecules, respectively. The shortened O-H distance of 1 Å directly indicates the deprotonation of $MA^+$ cation, only shown in +1 charge state, not in the other states (See **Fig. 4f**). Overall, intermolecular interactions between $MAPbI_3$ components and gaseous molecules actively take place when $MAPbI_3$ crystals are with localized hole, from which the fastest degradation of $MAPbI_3$ with localized hole can be clearly explained.

**Possible scenario on decrystallization of $MAPbI_3$ crystals under the mixture of $O_2$ and $H_2O$**

We suggest detailed scenario for degradation mechanisms of $MAPbI_3$ crystals under the mixture of $O_2$ and $H_2O$ from experiments and simulation results. In the present AIMD simulation, the mixture makes totally different intermolecular interactions from the situation when $O_2$ or $H_2O$ solely exist.[20] At the beginning state, $O_2$ molecule is easily adsorbed near the $I^-$ anion at the surface of $MAPbI_3$ regardless of charge state (see filled square line 0-80 fs of **Fig. 4e**), and it attempts to directly interact with Pb atoms (see 0-180 fs of **Fig. 4d**). However, different interplays of $H_2O$ and $O_2$ molecules with $MAPbI_3$ components occur depending on charge state (see **Fig. 4e-f**). For the case of +1 charge state (one localized hole), I-O interaction stabilizes (blue filled square line in **Fig. 4e**) and $H_2O$ molecule approaches closely to I-O (blue filled square line in **Fig. 4f**). Since positive charge of Pb atom is partially increased by a localized hole in the $MAPbI_3$ slab, Pb-O bond is finally formed after 500 fs (see **Fig. 4d**). Then, $H_2O$ molecule could completely provide a proton to Pb-bonded O atom, followed by forming stable hydroxide species (Pb-I-O-OH) (see blue filled square line in **Fig. 4f**). At the same time, MA cation is totally deprotonated by donating proton to deprotonated $H_2O$ ($OH^-$) (**Fig. 4f**). Consequently, such generation of hydroxide species and deprotonation of MA cation lead to fast destroy of a $MAPbI_3$ unit cell.

For the case of -1 charge state (one localized electron), I-O interaction abruptly weakens after 100 fs in **Fig. 4e**, which is likely to be influenced by augmented negative charge of I atom. Sequentially, Pb-O interaction is not strengthened enough to form stable Pb-O bond (see red line of **Fig. 4d**), and then $H_2O$ molecule separates from adsorbed $O_2$ molecule, which is indicative of no proton transfer (see red square line around 500 – 600 fs of **Fig. 4f**). In this case, $PbI_6^-$ octahedron is rather stabilized due to the absence of proton transfer. For the case of neutral charge state (no localized charge), no intermolecular interaction between gas molecules and components of $MAPbI_3$ occur, which means no degradation of $MAPbI_3$ crystal (see **Fig. 4b** and black lines of **Fig. 4d-f**). In summary, the decomposition of $MAPbI_3$ crystals under the mixed ambient of $O_2$ and $H_2O$ occurs vigorously through deprotonation with localized holes.

**X-ray assisted spectroscopic results supporting degradation scenario**

To find out experimental evidence of our degradation scenario, we measured and analyzed X-ray assisted spectroscopy data of fresh and degraded $MAPbI_3$ samples. First, we detected solid Pb(OH)I as one of lead hydroxide products appearing at the peak of 38.7° from X-ray diffraction patterns (XRD) of degraded $MAPbI_3$ films in **Fig. 5a**.[35] These results agree with the generation of lead hydroxide species in our AIMD simulation in the mixed ambient of $O_2$ and $H_2O$. Additionally, the existence of hydroxide was also observed from comparison of X-ray photoelectron spectroscopy (XPS) data for the fresh and degraded $MAPbI_3$ film. As shown in **Fig. 5b**, $O1_s$ peak shifts to higher binding energy region for the degraded sample, which means that O-H bonding character is reinforced.[36-38] In other words, it is certain that hydroxide species are produced as an end product of degradation as we supposed above. XPS spectra bring out another crucial evidence in terms of iodine loss. **Fig. 5c** shows XPS spectra of fresh $MAPbI_3$, the degraded ETL/$MAPbI_3$, and the degraded HTL/$MAPbI_3$ within the range from 135 eV to 140 eV. In the fresh sample, only Pb-I bond is detected as known from Pb $4f_{7/2}$ peak appearing at 138.0 eV. After the degradation, the distinct peak of 136.3 eV corresponding to metallic Pb appear in both cases of the ETL/MAPbI$_3$ and HTL/MAPbI$_3$ sample, which is indicative of iodine loss, namely the complete destruction of PbI$_6^-$ octahedral.[39-41] The peak intensity of the ETL/MAPbI$_3$ is much higher than that of the HTL/MAPbI$_3$, which is consistent with our simulation results in that MAPbI$_3$ crystal is more stable at the +1 charge state, as compared to the -1 charge state.

## Conclusions

We observed different light-induced degradation rates of $MAPbI_3$ films depending on the charge polarity selectivity of underlying transporting layer under various surrounding ambient conditions. It was confirmed that localized carrier polarity inside of $MAPbI_3$ film incurred by charge selectivity of the underlying layer plays a decisive role in determining the degradation rate of $MAPbI_3$ films. The $MAPbI_3$ films with localized hole degraded more rapidly in the presence of $H_2O$ than ones with localized electron, whereas inverted trend of degradation rate appeared under $O_2$-only ambient. Interestingly, the combination of $H_2O$ and $O_2$ led to the fastest degradation of the $MAPbI_3$ film only with localized hole, where the $MAPbI_3$ film with localized electron unexpectedly stabilized in contrast to the case of $O_2$-only condition. To explain these observations, we investigated intermolecular dynamics of $H_2O$, $O_2$ and components of $MAPbI_3$ through DFT-based AIMD simulations. The dramatic difference in degradation rates under mixed atmosphere condition is mainly attributed to the generation of strong Pb-O interaction and lead hydroxide species at the hole localized $MAPbI_3$ crystal, which was evidenced by XPS and XRD results. Our study suggests that the positive localized charge in the $MAPbI_3$ crystals drives fast de-crystallization under actual atmospheric conditions, which provide the direction for long-term stability of perovskite solar cells.

## Methods

**Materials and device fabrication**

All materials were used as received. The ITO glass substrates (AMG, 9.5 Ω cm$^{-2}$) were cleaned by sonication in acetone, isopropanol, and deionized water followed by the deposition of the transporting materials. A 35 nm thick C$_{60}$ layer was coated onto the ITO glass substrate using a vacuum thermal evaporator at a deposition rate of 0.2 Å·s$^{-1}$ and pressure of 10$^{-7}$ Torr. A single PCBM layer was spin-coated at a rotation speed of 2,000 rpm. for 60 s using its 10 mg/mL solution in chlorobenzene (Sigma-Aldrich) and then annealed at a temperature of 100 °C for 10 min. A Spiro-MeOTAD (Merk) layer was drop-cast on a spinning substrate at rotation speed of 3,000 rpm for 30 s using its 72.3 mg/mL solution in chlorobenzene. A PEDOT: PSS (Clevios, AI4083) layer was spin-coated at a speed of 5,000 rpm for 40 s using its 1:1 (v/v) diluted solution in deionized water and then annealed at 130 °C for a minimum of 30 min. MAPbI$_3$ perovskite solutions were prepared by mixing PbI$_2$ (Alfa Aesar), MAI (Xian' Chemical), dimethyl sulfoxide (DMSO, Sigma-Aldrich), and dimethylformamide (DMF, Sigma-Aldrich) at specified proportions. For example, to prepare 52 wt.% of precursor in DMF solvent with the molecular ratio PbI$_2$:MAI:DMSO = 1:1:1, 461 mg of PbI$_2$, 159 mg of MAI, and 78 mg of dimethyl sulfoxide were mixed in 0.60 mL of DMF. After spin coating the obtained precursor at a speed of 4,000 rpm for 20 s combined with a diethyl ether dripping procedure, the resulting intermediate films were annealed at 100 °C for 10 min. All spin coating procedures in this study were performed inside a dry room with a relative humidity of less than 10% and temperature of 25 °C.

**Ageing conditions**

All perovskite films were exposed to light in a controlled environment by customized chamber. Dry air condition was configured by supplying dry air gas (80 % nitrogen / 20 % oxygen) as 0.5 liter/m. $H_2O$ condition was configured by supplying nitrogen gas with $H_2O$ bubbler as 0.5 liter/m. The mixed ambient condition was configured by supplying dry air gas with $H_2O$ bubbler as 0.5 liter/m. Light illumination was performed under 1 sun condition using an AAA solar simulator (Newport Oriel Solar 3A Class AAA, 64023A). The light intensity was calibrated using a KG–5 standard silicon solar cell (Oriel, VLSI Standards).

**Characterization**

Time-resolved and steady-state PL measurements were conducted using a FluoroMax-4 spectrofluorometer (Horiba). The studied films were photo-excited with a 463-nm laser diode (DeltaDiode-470L, Horiba) pulsed at a frequency of 100 MHz. The resulting PL was detected by a high-sensitivity photon counting near infrared (NIR) detector. Planar SEM and cross-sectional images of the perovskite solar cell structure were obtained using a field-emission scanning electron microscope (FESEM, AURIGA, Carl Zeiss) combined with a focused ion beam system (FIB system, AURIGA, Carl Zeiss). The absorbance spectra and crystalline diffraction patterns were recorded using an ultraviolet-visible/NIR spectrophotometer (Cary 5000, Agilent Technologies) and an X-ray diffractometer (New D8 Advanced, Bruker), respectively. The XPS spectra were obtained using an electron spectroscopy for chemical analysis instrument (Sigma Probe, VG Systems). SKPM measurements was performed in a non-contact mode using an atomic force microscope (NX10, Park Systems) equipped with an NSC36/Cr-Au tip.

**DFT calculation**

The DFT calculations were conducted using the projector augmented wave (PAW) method and generalized gradient approximation of Perdew, Burke, and Ernzerhof (PBE) for the exchange-correlation potential implemented in the Vienna Ab-initio Simulation Package (VASP) code. The lattice parameters a and c of the tetragonal $(CH_3NH_3)PbI_3$ lattice were 8.86 Å and 12.66 Å for 4 unit cells, respectively. Monkhorst-Pack k-point sampling with 4×4×2 k-point grids was used for the Brillouin zone integration. An energy cutoff of 500 eV was used to expand the wavefunctions in the plane-wave representation. Atomic structures were relaxed until all Hellman-Feynman forces were below 0.01 eV/Å. A climbing-image nudged elastic band method was used for locating minimum energy pathways.

## Acknowledgements

This work has been supported by the Global Frontier R&D Program of the Center for Multiscale Energy Systems funded by the National Research Foundation under the Ministry of Science and ICT, Korea (2012M3A6A7054855). This work has been also supported by the National Research Foundation under the Ministry of Science and ICT, Korea. (2017R1A2B3010927). The XRD patterns were measured at Research Institute of Advanced Materials in Seoul National University.

## Author contributions

M.-c.K., N.A., Y.U.J., H.S.J. and M.C. conceived and designed the experiments. M.-c.K. performed the half device fabrication and carried out the controlled stability test. M.-c.K. and N. A. measured KPFM. M.-c.K., N.A., and P.V.P discussed the mechanism of charge unbalance generation in $MAPbI_3$. E.L and S.K.K calculated the DFT simulations. M.-c.K., N.A., H.S.J. and M.C. wrote the paper. All authors discussed the results.

2015, **3**, 9081-9085.

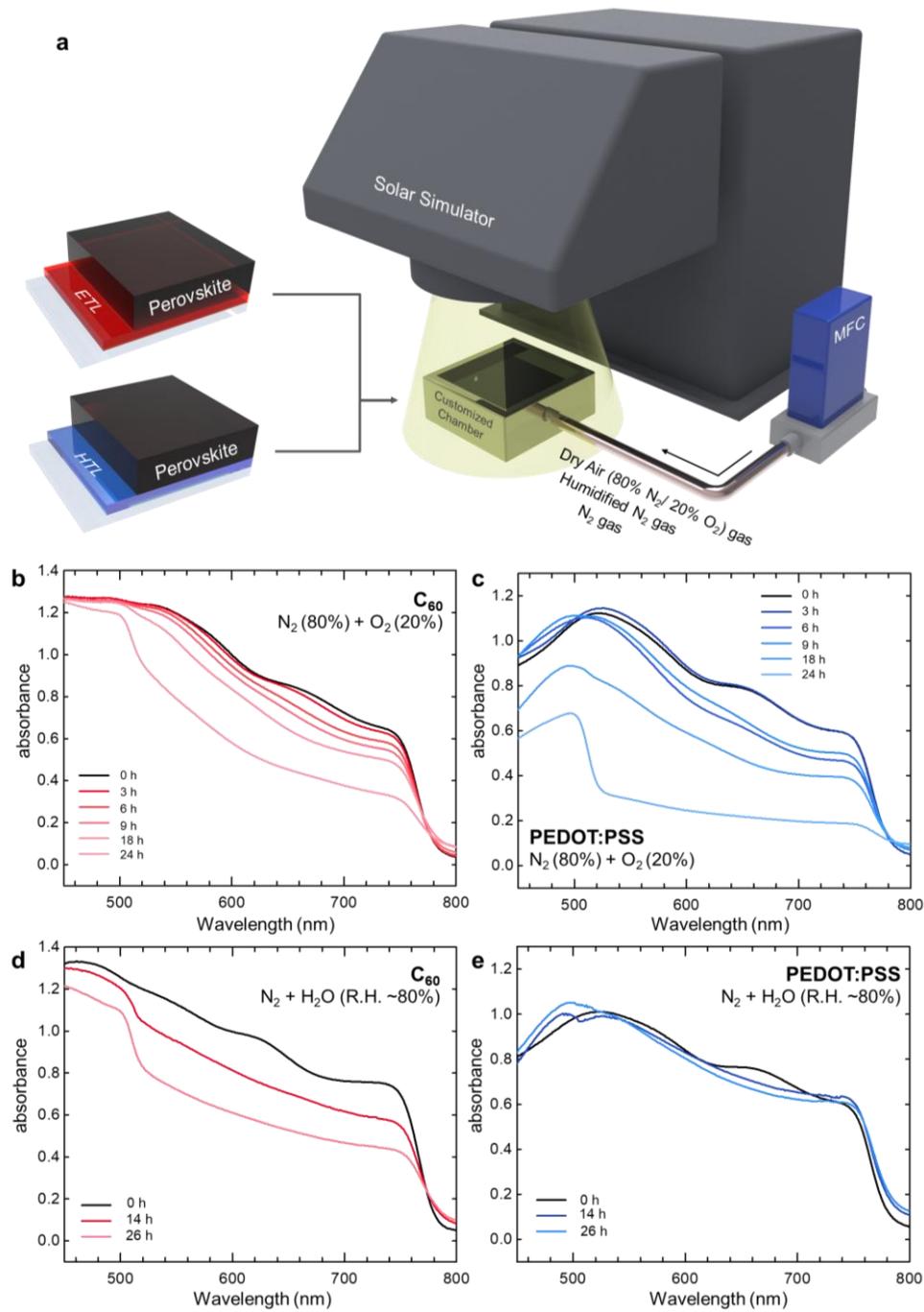

**Fig. 1. Light-induced degradation test of perovskite films depending on the excessive charge polarity under different atmospheric conditions.** (a) Schematics of customized experimental systems for light-induced degradation with solar simulator, customized chamber and gas flow controller. Time evolution of absorption spectra for (b) the $C_{60}$/MAPbI$_3$ (hole-rich) and (c) the PEDOT:PSS/MAPbI$_3$ (electron-rich) half devices in a chamber filled with dry air (80% nitrogen and 20% oxygen). Time evolution of absorption spectra for (d) the $C_{60}$/MAPbI$_3$ and (e) the PEDOT:PSS/MAPbI$_3$ half devices in a chamber filled with nitrogen and water vapors (R.H. ~ 80%). Every samples were under continuous light illumination except when measuring absorption spectra.

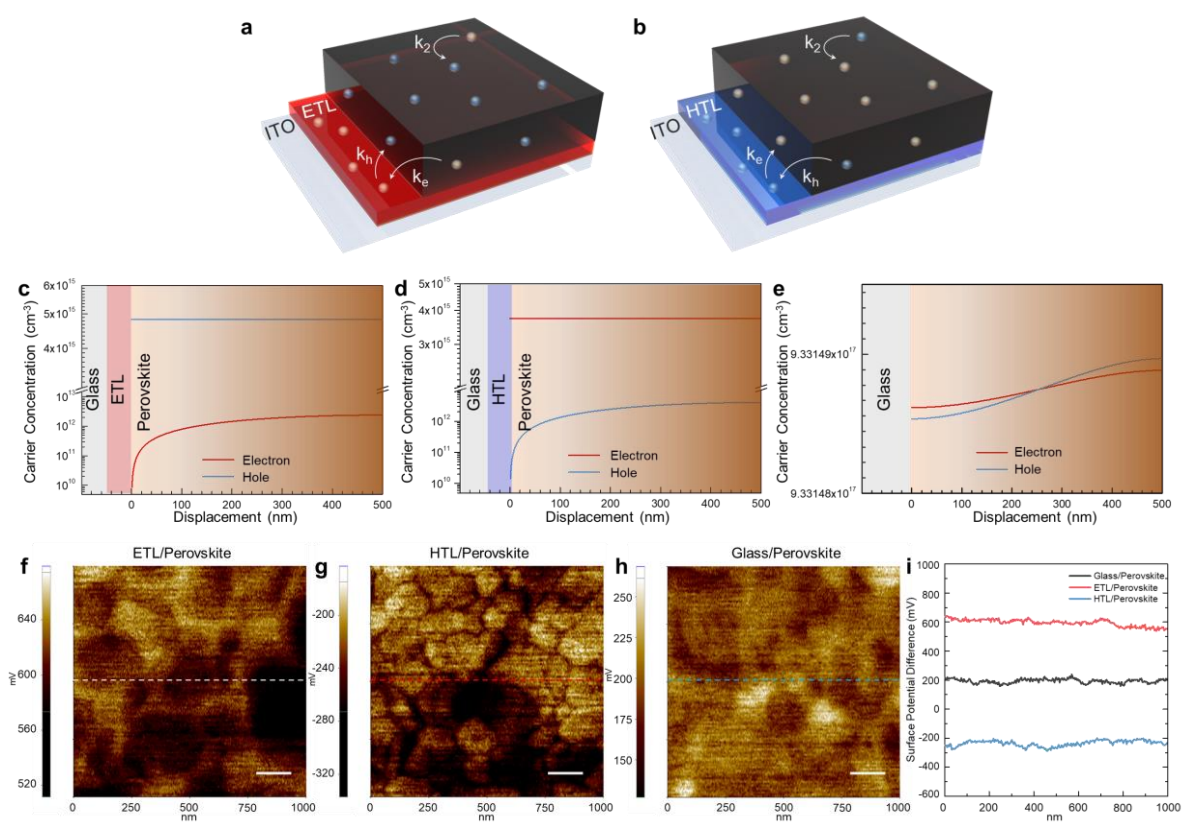

**Fig. 2. Unbalance of charge polarity for chare-selective transporting half devices.** Schematics of the experimental setup and charge-selective transporting half devices ((a)ETL/MAPbI$_3$ and (b)HTL/MAPbI$_3$ ones). Calculated carrier concentration as a function of thickness displacement for MAPbI$_3$ film deposited on (c) Glass/ETL, (d) Glass/HTL and (e) Glass substrate. Surface potential energy distribution images measured by scanning Kelvin probe microscopy (SKPM) for MAPbI$_3$ film deposited on (f) C$_{60}$ (ETL), (g) PEDOT:PSS (HTL) and (h) glass substrate. (i) Surface potential plot along the marked line in (f-h) respectively. All the scale bars are 150 nm.

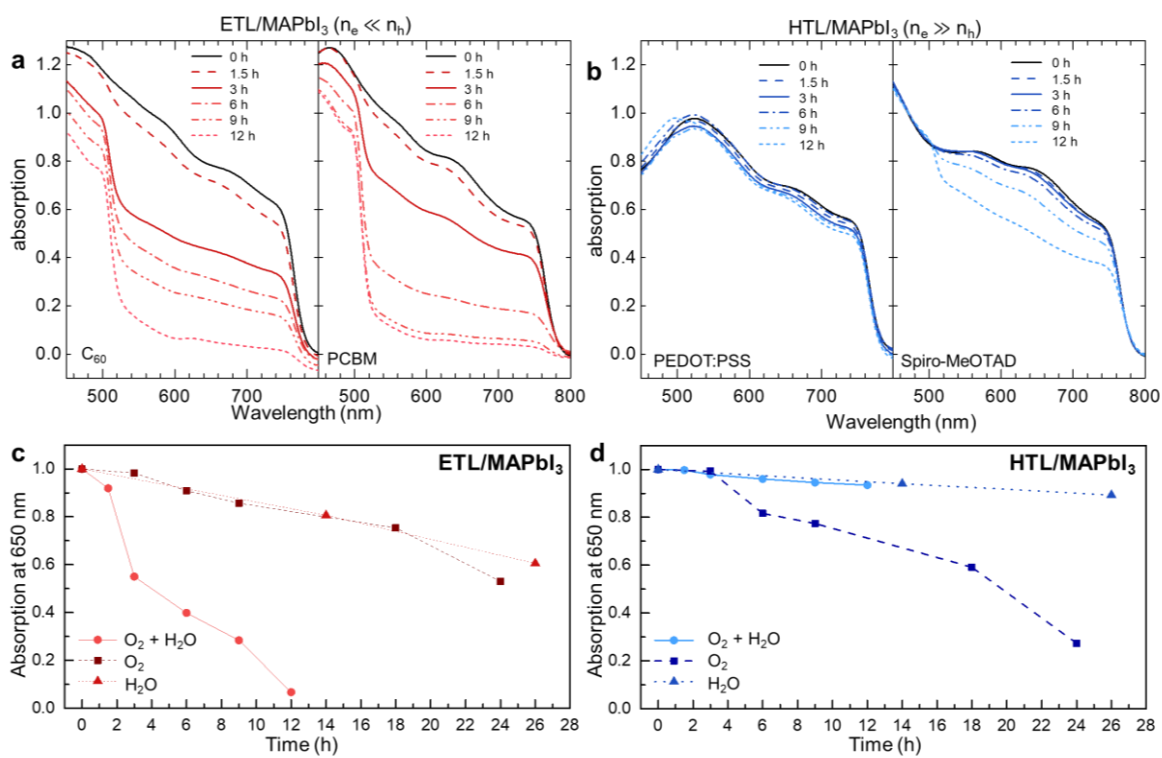

**Fig. 3. Light-induced degradation under the mixture ambient of $O_2$ and $H_2O$ depending on localized charge polarity.** Time evolution of absorption spectra from 0 h to 12 h for (a) ETL/MAPbI$_3$ and (b) HTL/MAPbI$_3$ samples. $C_{60}$, PCBM were used for ETL/MAPbI$_3$ configuration and PEDOT:PSS, Spiro-MeOTAD for HTL/MAPbI$_3$ configuration. Time evolution of absorption spectra are summarized by plotting absorption peak at 650 nm for (c) ETL/MAPbI$_3$ and (d) HTL/MAPbI$_3$ under 3 different ambient conditions (the mixture ambient, $O_2$ (20%) ambient and $H_2O$ (R.H. ~80%) ambient).

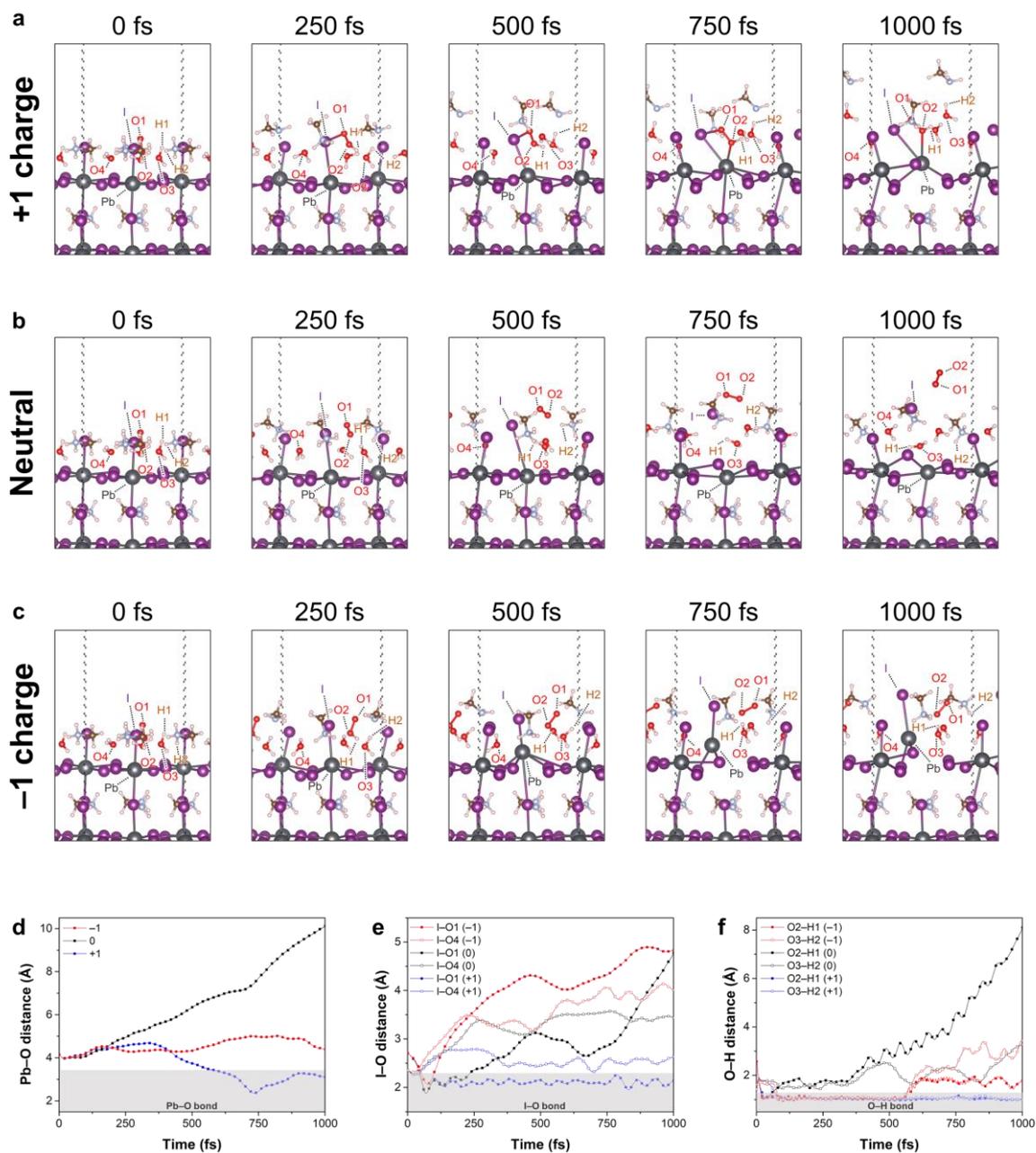

**Fig. 4. AIMD simulation results.** Time evolution of molecular dynamic simulations of MAPbI$_3$ surface for (a) +1 unit charge, (b) neutral, (c) -1 unit charge under 3 H$_2$O and 1 O$_2$ ambient atmosphere. Atomic bonding distance of (d) Pb-O, (e) I-O, (f) O-H bond respectively.

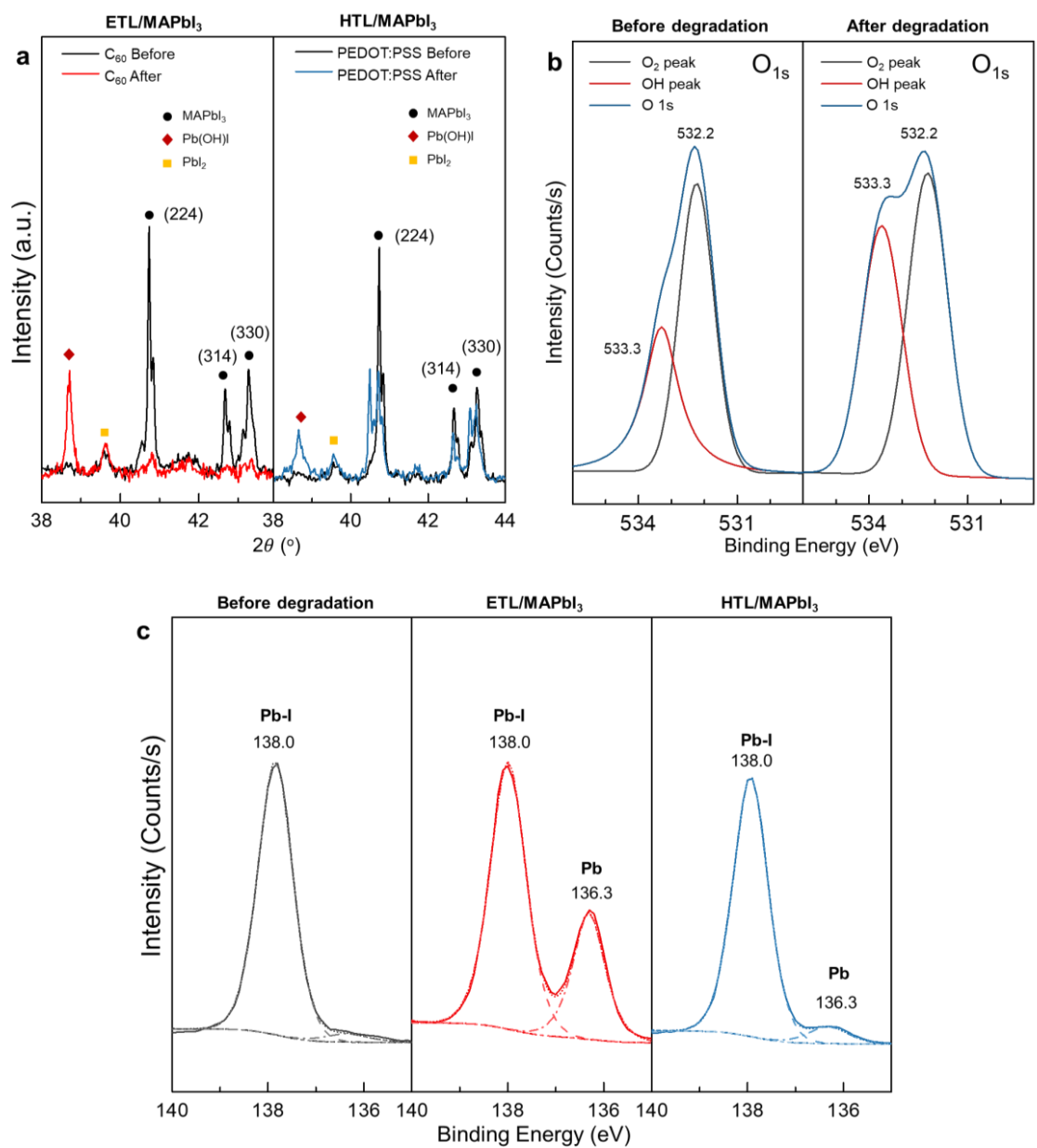

**Fig. 5. Observation of hydroxide species and Metallic Pb as the result of light-induced perovskite film degradation.** (a) XRD patterns of degraded MAPbI$_3$ films deposited on ETL and HTL at the range between 38º to 44º 2$\theta$. MAPbI$_3$, Pb(OH)I and PbI$_2$ peaks are indexed respectively. (b) XPS spectra recorded for the O atom (O1s) of the ETL/MAPbI$_3$ film before and after degradation (c) XPS spectra recorded for the Pb atom (Pb 4f$_{7/2}$) of the fresh, ETL/MAPbI$_3$ and HTL/MAPbI$_3$ samples.

Electronic Supplementary Information

# Decrystallization of CH$_3$NH$_3$PbI$_3$ perovskite crystals via polarity dependent localized charges

*Min-cheol Kim$^{a\dagger}$, Namyoung Ahn$^{a\dagger}$, Eunhak Lim$^b$, Young Un Jin$^c$, Peter V. Pikhitsa$^a$, Jiyoung Heo$^d$, Seong Keun Kim$^b$, Hyun Suk Jung$^{c*}$ and Mansoo Choi$^{a,e*}$*

$^a$Global Frontier Center for Mulitscale Energy Systems, Seoul National University, Seoul, Republic of Korea.

$^b$Department of Chemistry, Seoul National University, Seoul, Republic of Korea.

$^c$School of Advanced Materials Science & Engineering, Sungkyunkwan University, Suwon, Gyeonggi-do, Republic of Korea.

$^d$Department of Green Chemical Engineering, Sangmyung University, Seoul, Republic of Korea.

$^e$Department of Mechanical Engineering, Seoul National University, Seoul, Republic of Korea.

$^\dagger$These authors contributed equally to this work.

$^*$Correspondence and request for materials should be addressed to H.S.J, and M. C.

(email: hsjung1@skku.edu; mchoi@snu.ac.kr)

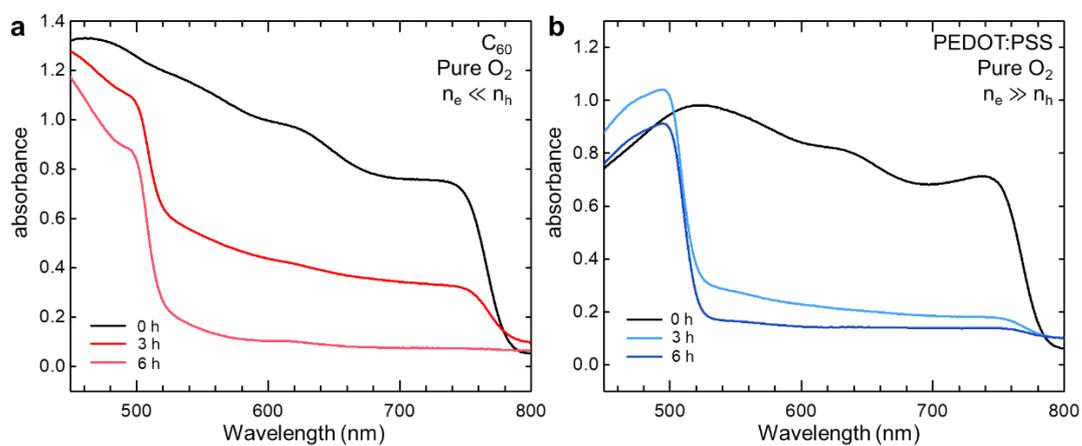

**Supplementary Figure 1.** Time evolution of absorption spectra for (a) the $C_{60}$/MAPbI$_3$ and (b) the PEDOT:PSS/MAPbI$_3$ half devices in chamber filled with pure $O_2$ gas before and after light-illumination for 6 h.

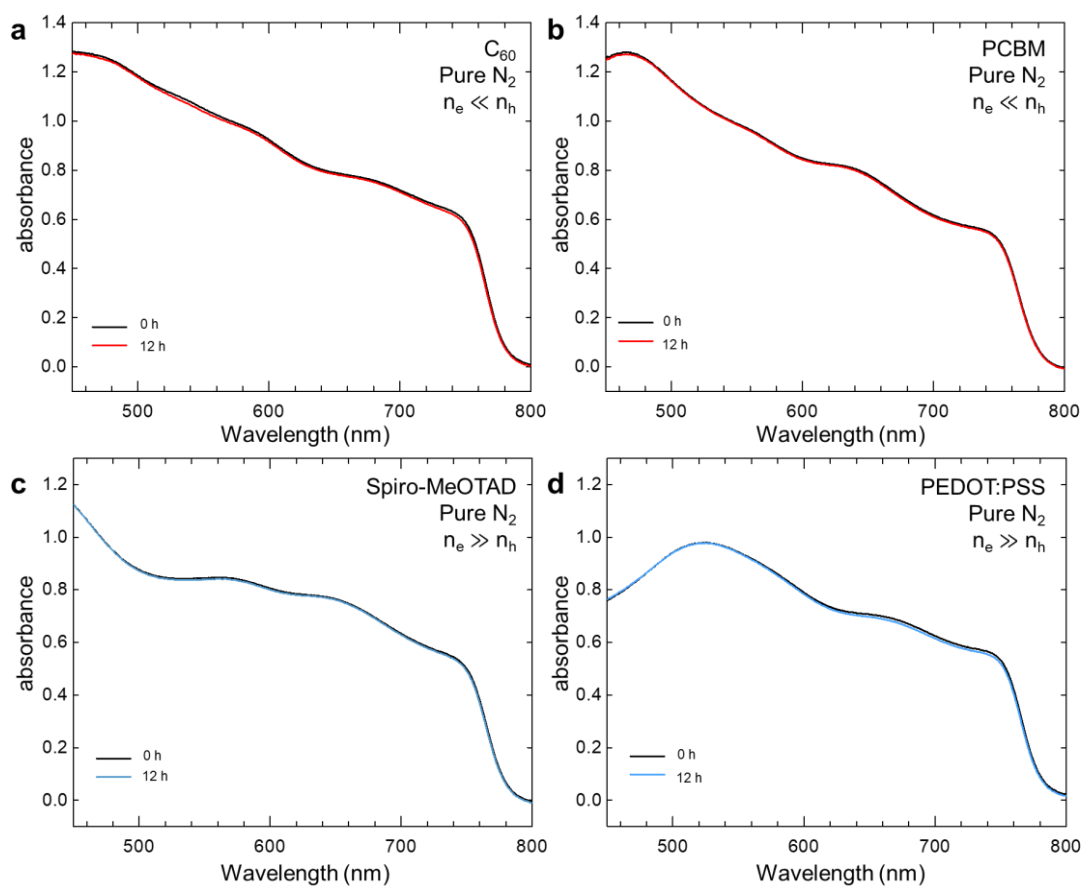

**Supplementary Figure 2.** Absorption spectra of perovskite films deposited on (a) $C_{60}$, (b) PCBM, and perovskite films deposited on (c) Spiro-MeOTAD and (d) PEDOT:PSS in chamber filled with pure $N_2$ gas before and after light-illumination for 12 h.

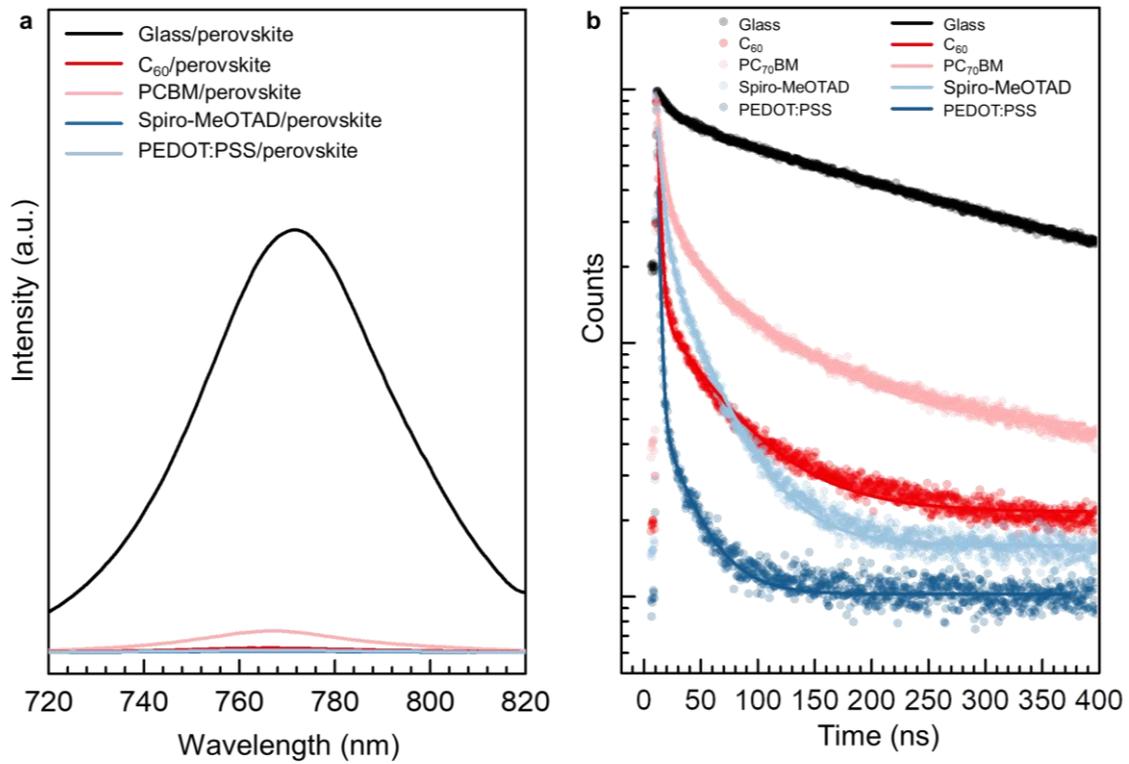

**Supplementary Figure 3.** (a) Steady-state and (b) time-resolved PL spectra obtained for the perovskite films deposited on the glass, PCBM, $C_{60}$, Spiro-MeOTAD, and PEDOT:PSS substrates and fitted with the three-term exponential decay function. Time-resolved PL spectra is plotted as circle, and fitted function is plotted as line.

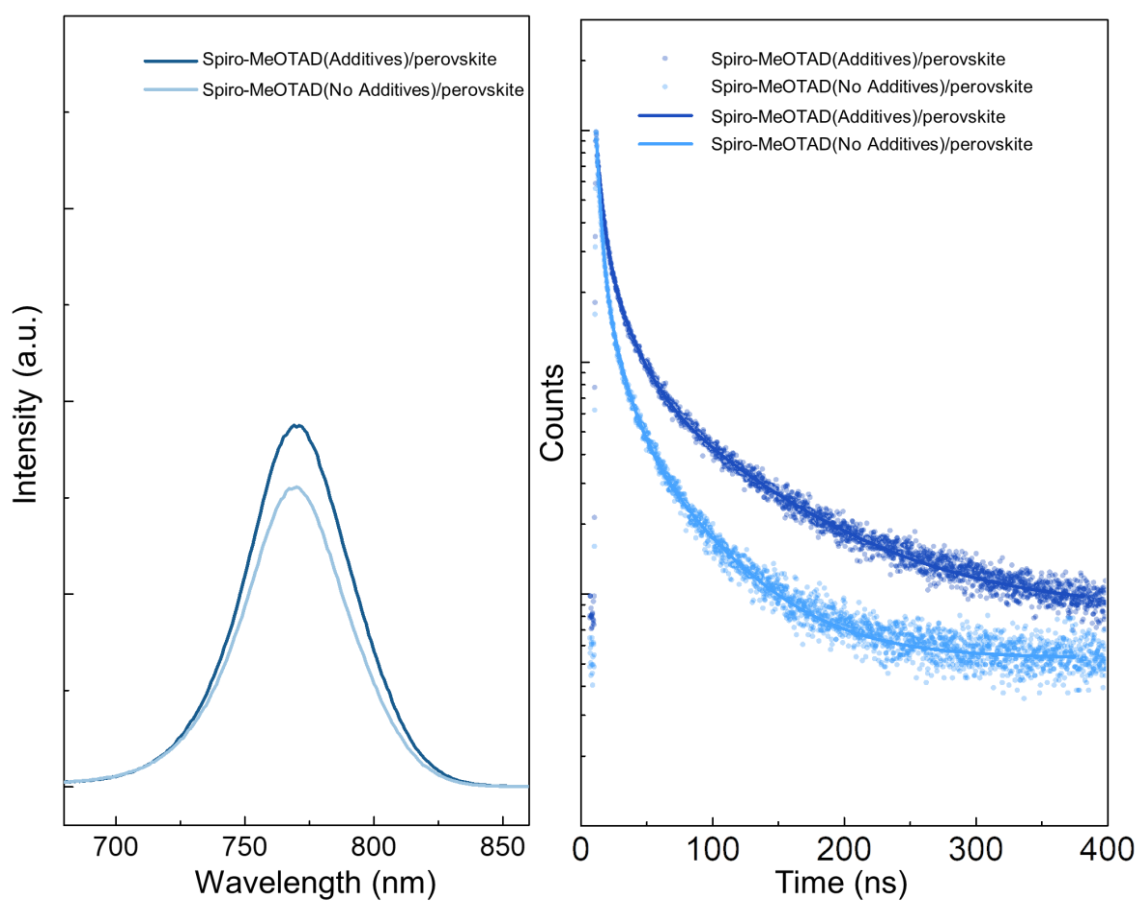

**Supplementary Figure 4.** (a) Steady-state and (b) time-resolved PL spectra obtained for the perovskite films deposited on the Spiro-MeOTAD with and without additives (such as tBP and Li-TFSI). The data fitted with the three-term exponential decay function.

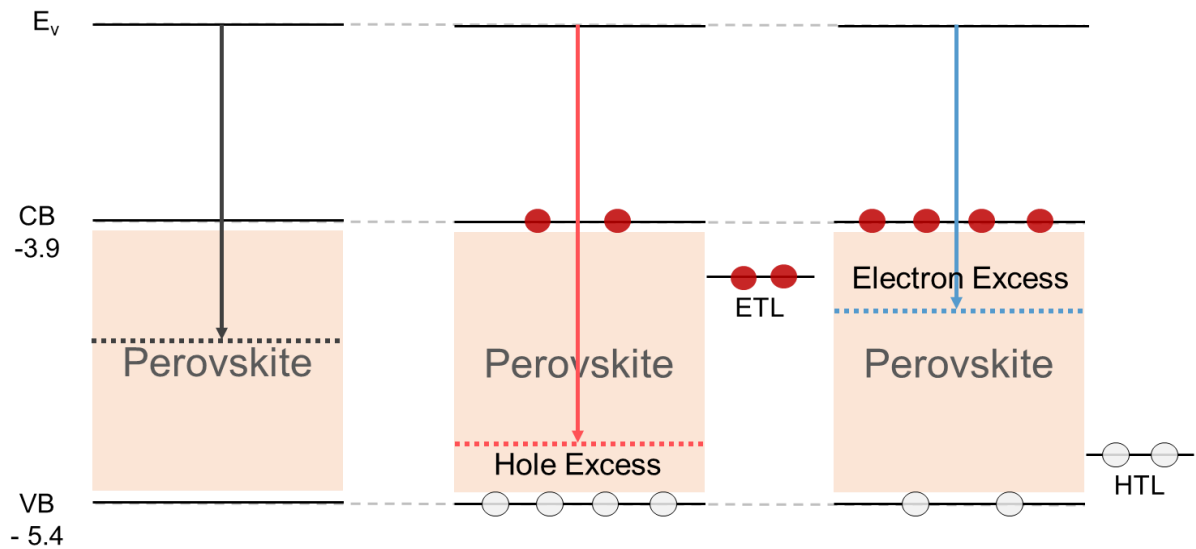

**Supplementary Figure 5.** Band energy diagram for perovskite films deposited on (a) glass, (b) ETL, and (c) HTL. Quasi fermi level of each films altered due to imbalance of electron-hole carrier concentration. Hole-excess perovskite film deposited on ETL presented lower quasi fermi level, therefore higher surface potential (work function). On the other hands, electron-excess perovskite film on HTL showed lower surface potential.

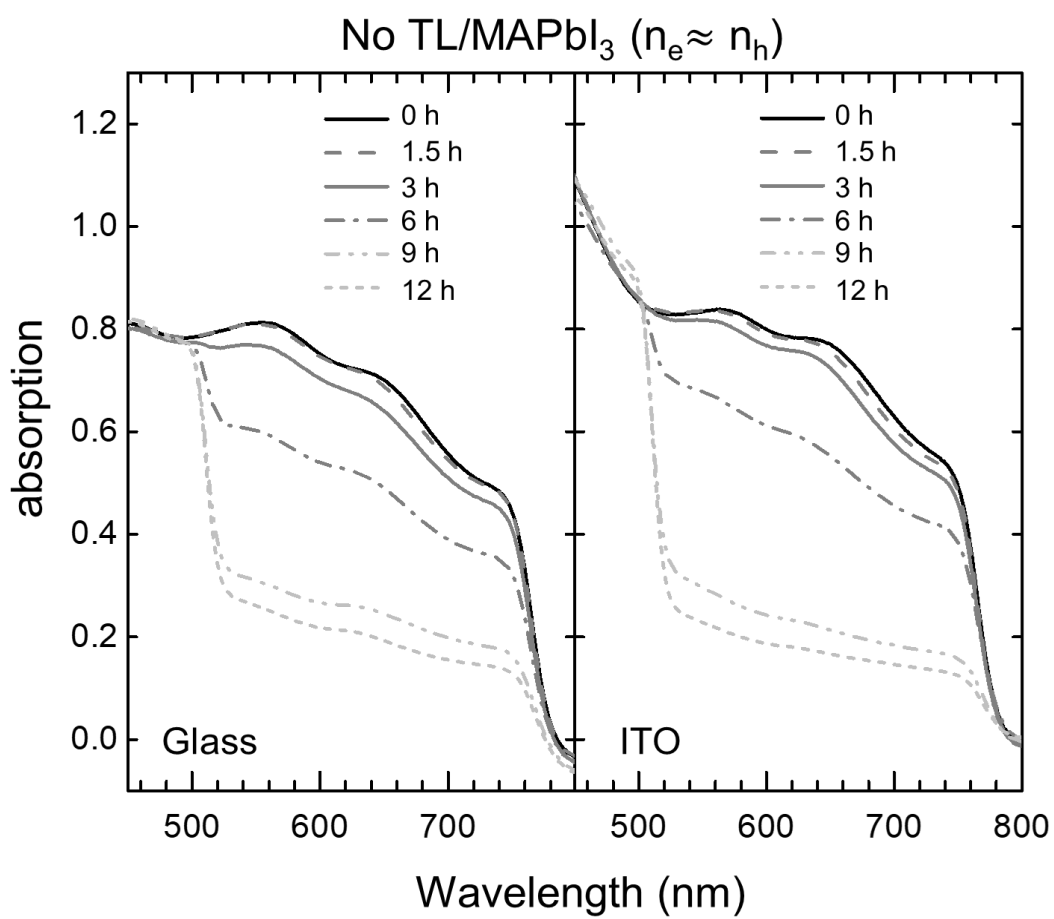

**Supplementary Figure 6.** (a) Steady-state and (b) time-resolved PL spectra obtained for the perovskite layers deposited on the Spiro-MeOTAD with and without additives (such as tBP and Li-TFSI). The data fitted with the three-term exponential decay function.

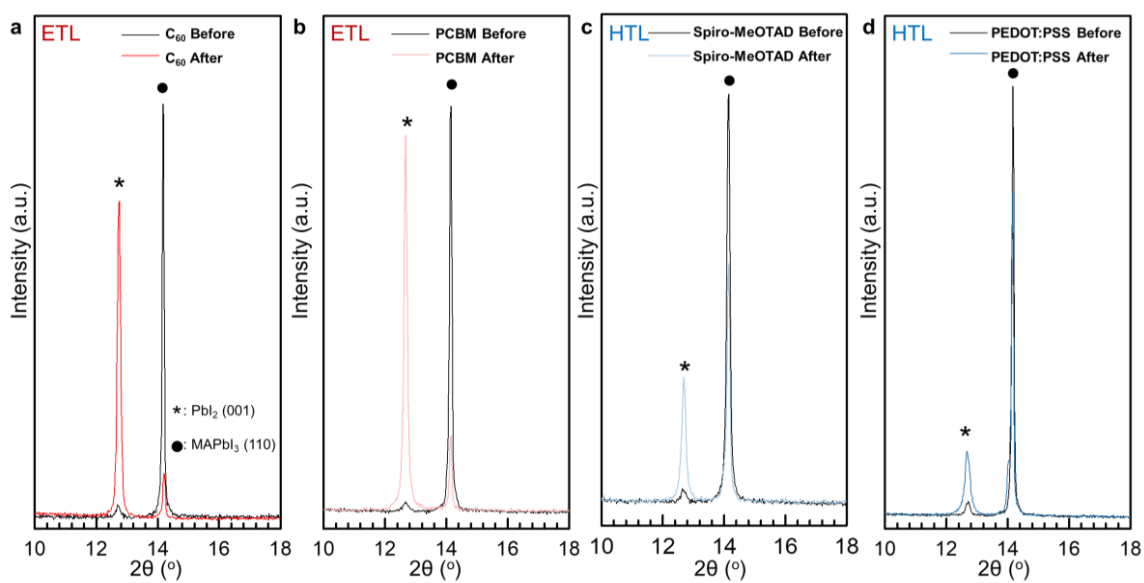

**Supplementary Figure 7.** XRD reflections of the (a) $C_{60}$, (b) PCBM, (c) Spiro-MeOTAD, and (d) PEDOT:PSS phases in the 2θ range of 10–18°. $PbI_2$ species are detected at 12.7°, and $MAPbI_3$ species are detected at 14.2°.

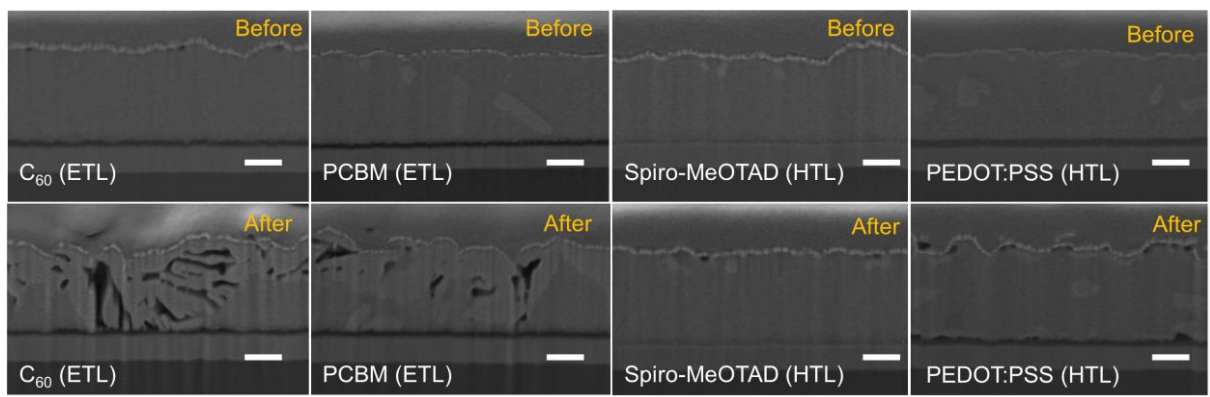

**Supplementary Figure 8.** The cross-sectional SEM images of the (e) $C_{60}$, (f) PCBM, (g) Spiro-MeOTAD, and (h) PEDOT:PSS layers obtained by using a focused ion beam before and after light exposure for 3 h. All scale bars are equal to 200 nm.

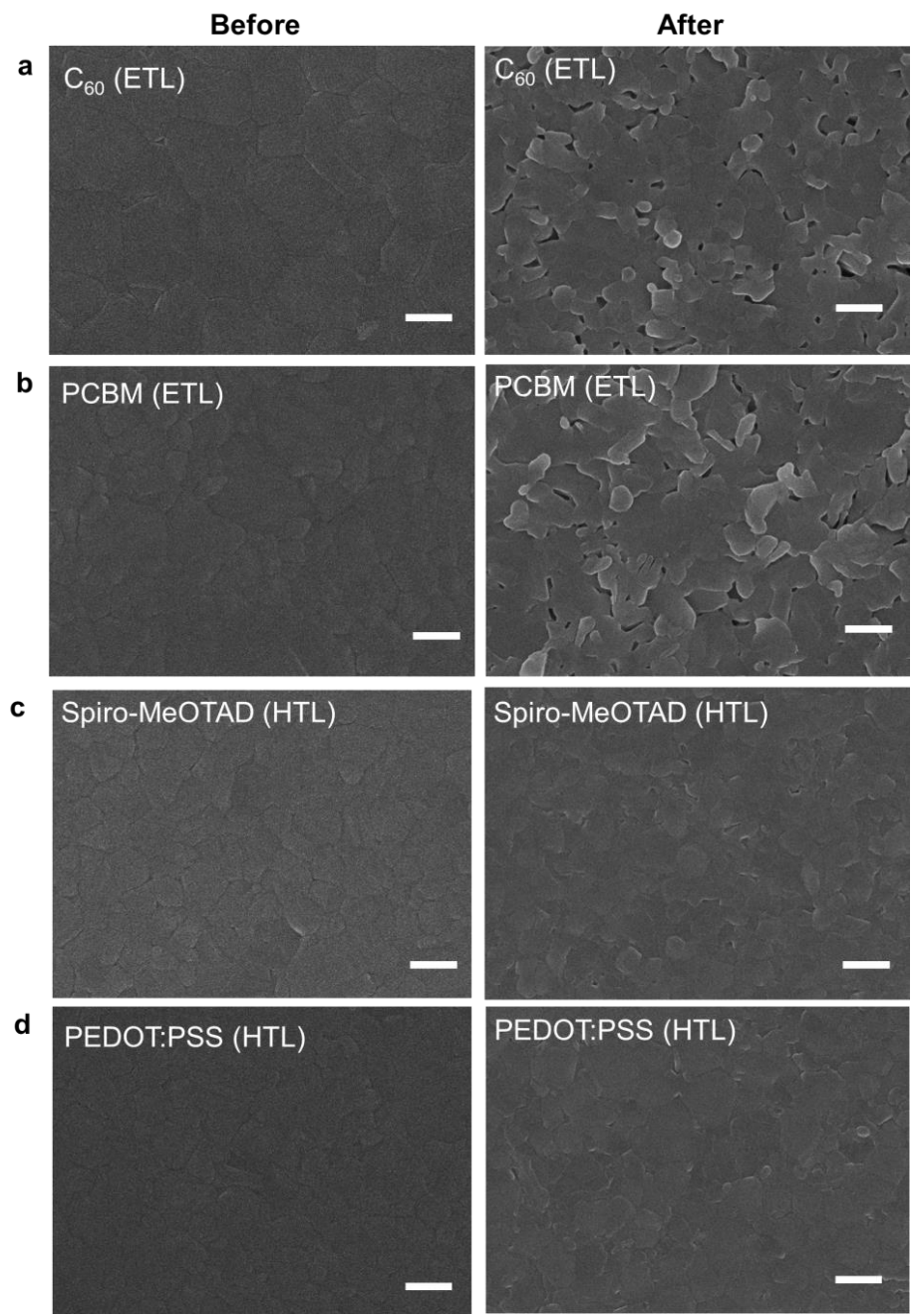

**Supplementary Figure 9.** Plane-view SEM images of the perovskite layers deposited on the (a) $C_{60}$, (b) PCBM, (c) Spiro-MeOTAD, and (d) PEDOT:PSS substrates before and after 6 h of light-soaking. All scale bars are equal to 300 nm.

|                | A$_1$(%) | τ$_1$ (ns) | A$_2$(%) | τ$_2$ (ns) | A$_3$(%) | τ$_3$ (ns) | Avg. τ |
|----------------|----------|------------|----------|------------|----------|------------|--------|
| **Glass**      | 0.36     | 7.87       | 1.64     | 31.95      | 98.00    | 332.89     | 332.38 |
| **C$_{60}$**   | 13.93    | 1.99       | 26.65    | 22.02      | 59.42    | 103.82     | 96.33  |
| **PCBM**       | 6.80     | 3.86       | 24.24    | 28.96      | 68.96    | 151.07     | 143.03 |
| **PEDOT:PSS**  | 12.11    | 3.76       | 48.01    | 20.99      | 48.01    | 52.69      | 43.15  |
| **Spiro-MeOTAD**| 25.88   | 1.47       | 20.82    | 5.17       | 53.30    | 32.97      | 30.77  |

**Supplementary Table 1.** A summary of the time-resolved PL lifetime parameters obtained for the glass, C$_{60}$, PCBM, PEDO:PSS, and Spiro-MeOTAD/perovskite systems described in Supplementary Figure 2b. Intensity weighted average lifetime was calculated by $\tau_{avg} = \frac{\sum_i A_i \tau_i^2}{\sum_i A_i \tau_i}$

| Bottom Layer | | PbI$_2$ (12.7°) | MAPbI$_3$ (14.2°) |
|---|---|---|---|
| C$_{60}$ (ETL) | Before | 428 | 7867 |
| | After | 6077 | 1031 |
| | Ratio | 18.38 | 17.0% |
| PCBM (ETL) | Before | 632 | 9895 |
| | After | 9215 | 2181 |
| | Ratio | 14.58 | 22.0% |
| Spiro-MeOTAD (HTL) | Before | 651 | 7599 |
| | After | 2614 | 4612 |
| | Ratio | 4.02 | 60.7% |
| PEDOT:PSS (HTL) | Before | 772 | 13804 |
| | After | 2371 | 10508 |
| | Ratio | 3.07 | 76.1% |

**Supplementary Table 2.** XRD intensities of the PbI$_2$ (12.7°) and MAPbI$_3$ (14.2°) peaks obtained for the perovskite layers deposited on the C$_{60}$, PCBM, Spiro-MeOTAD, and PEDOT:PSS substrates before and after light exposure.

Electronic Supplementary Information

# Decrystallization of CH$_3$NH$_3$PbI$_3$ perovskite crystals via polarity dependent localized charges

*Min-cheol Kim[a†], Namyoung Ahn[a†], Eunhak Lim[b], Young Un Jin[c], Peter V. Pikhitsa[a], Jiyoung Heo[d], Seong Keun Kim[b], Hyun Suk Jung[c*] and Mansoo Choi[a,e*]*

[a]Global Frontier Center for Mulitscale Energy Systems, Seoul National University, Seoul, Republic of Korea.

[b]Department of Chemistry, Seoul National University, Seoul, Republic of Korea.

[c]School of Advanced Materials Science & Engineering, Sungkyunkwan University, Suwon, Gyeonggi-do, Republic of Korea.

[d]Department of Green Chemical Engineering, Sangmyung University, Seoul, Republic of Korea.

[e]Department of Mechanical Engineering, Seoul National University, Seoul, Republic of Korea.

[†]These authors contributed equally to this work.

[*]Correspondence and request for materials should be addressed to H.S.J, and M. C.

(email: hsjung1@skku.edu; mchoi@snu.ac.kr)

To estimate electron and hole densities inside MAPbI$_3$ layer, the structure is simplified as shown in Supplementary Fig.1. All the layers are ideally uniform along y-direction for building 1D simple model. Thus, we considered carrier densities as a function of thickness displacement along x-direction. Additionally, we assumed that there is no ion migration, no trap-assisted recombination and no photon recycling because those are not dominant under one sun illumination conditions (mainly generation and dynamics of charge carriers). Consequently, we considered light absorption (photo-carrier generation), radiative recombination, carrier diffusion and carrier drift inside MAPbI$_3$ perovskite layers by utilizing its physical properties according to the references [ref].

First, the carrier transport model is regarded as the following Boltzmann transport equations.

$$J_e = -q\mu_e n_e \nabla \varphi_{F,e} \qquad \text{Eq 1}$$

$$J_h = -q\mu_h n_h \nabla \varphi_{F,h} \qquad \text{Eq 2}$$

Where q is the elementary charge, $\mu_e$ and $\mu_h$ are the mobility of electron and hole, $n_e$ and $n_h$ are the density of electron and hole, $\varphi_{F,e}$ and $\varphi_{F,h}$ are the quasi Fermi level of electron and hole, respectively.

The quasi Fermi level is given by

$$\varphi_{F,e} = E_{CB} + V - \frac{kT}{q} \ln \frac{n_e}{N_0} \qquad \text{Eq 3}$$

$$\varphi_{F,h} = E_{VB} + V + \frac{kT}{q} \ln \frac{n_h}{N_0} \qquad \text{Eq 4}$$

Where $E_{cb}$ and $E_{vb}$ are the energy level of conduction band and valence band, V is the electrostatic potential, k is Boltzmann's constant, T is the temperature, $N_0$ is the total density

of state

Accordingly, the current density of electrons and holes ($J_e$ and $J_h$) could be calculated as follows.

$$\therefore J_e = -q\mu_e n_e \left(\frac{dV}{dx} - \frac{kT}{q}\frac{1}{n_e}\frac{dn_e}{dx}\right) = -q\mu_e n_e \frac{dV}{dx} + \mu_e kT \frac{dn_e}{dx}$$
... Eq 5

$$\therefore J_h = -q\mu_h n_h \left(\frac{dV}{dx} + \frac{kT}{q}\frac{1}{n_h}\frac{dn_h}{dx}\right) = -q\mu_h n_h \frac{dV}{dx} - \mu_h kT \frac{dn_h}{dx}$$
... Eq 6

The first term is associated with carrier drift, and the second one is corresponding to carrier diffusion by density difference.

To calculate carrier densities, the continuity equation of carriers is required. Considering carrier generation (G) and carrier recombination (R), the continuity equation of electrons and holes are expressed as the following equations.

$$\frac{dn_e}{dt} = \frac{1}{q}\frac{dJ_e}{dx} + G - R \qquad \text{Eq 7}$$

$$\frac{dn_h}{dt} = -\frac{1}{q}\frac{dJ_h}{dx} + G - R \qquad \text{Eq 8}$$

Recombination R is calculated by following a Langevin process as expressed in Eq. 9, and generation G is calculated according to the reference.[1]

$$R = L\frac{q(\mu_e + \mu_h)}{\varepsilon_0 \varepsilon_r} n_e n_h$$
Eq 9

Finally, we can obtain the continuity equation of carrier in the 1D model by using Eq, 5-9, as shown in below.

$$\frac{dn_h}{dt} = -\frac{1}{q}\frac{dJ_h}{dx} + G - R = \mu_h n_h \frac{d^2V}{dx^2} + \mu_h \frac{dn_h}{dx}\frac{dV}{dx} + \frac{\mu_h kT}{q}\frac{d^2n_h}{dx^2} - L\frac{q(\mu_e + \mu_h)}{\varepsilon_0\varepsilon_r}n_e n_h + G$$

Eq 10

$$\frac{dn_e}{dt} = \frac{1}{q}\frac{dJ_e}{dx} + G - R = -\mu_e n_e \frac{d^2V}{dx^2} - \mu_e \frac{dn_e}{dx}\frac{dV}{dx} + \frac{\mu_e kT}{q}\frac{d^2n_e}{dx^2} - L\frac{q(\mu_e + \mu_h)}{\varepsilon_0\varepsilon_r}n_e n_h + G$$

Eq 11

At the steady-state, the time derivative of carrier density is zero ($dn_e/dt = dn_h/dt = 0$). Therefore, Eq 10 and 11 could be simplified more.

$$\frac{\mu_e kT}{q}\frac{d^2n_e}{dx^2} = \mu_e n_e \frac{d^2V}{dx^2} + \mu_e \frac{dn_e}{dx}\frac{dV}{dx} + L\frac{q(\mu_e + \mu_h)}{\varepsilon_0\varepsilon_r}n_e n_h - G$$

Eq 12

$$\frac{\mu_h kT}{q}\frac{d^2n_h}{dx^2} = -\mu_h n_h \frac{d^2V}{dx^2} - \mu_h \frac{dn_h}{dx}\frac{dV}{dx} + L\frac{q(\mu_e + \mu_h)}{\varepsilon_0\varepsilon_r}n_e n_h - G$$

Eq 13

Potential terms can be solved from poisson equation of Gauss law.

$$\nabla^2 V = -\frac{q}{\varepsilon_0\varepsilon_r}(n_h - n_e) \qquad \text{Poisson equation}$$

$$\frac{d^2V}{dx^2} = -\frac{q}{\varepsilon_0\varepsilon_r}(n_h - n_e) \qquad \text{Poisson equation in 1D model}$$

Finally, the continuity equation of electrons and holes at the steady-state are obtained as a function of electron and hole densities and thickness displacement x as the following equations.

$$\therefore \frac{\mu_e kT}{q}\frac{d^2n_e}{dx^2} = \frac{q\mu_e}{\varepsilon_0\varepsilon_r}(n_e^2 - n_e n_h) + \mu_e \frac{dn_e}{dx}\frac{q}{\varepsilon_0\varepsilon_r}\left(\int (n_e - n_h)dx - E_{x0}\right) + L\frac{q(\mu_e + \mu_h)}{\varepsilon_0\varepsilon_r}n_e n_h - G$$

Eq 14

$$\therefore \frac{\mu_h kT}{q}\frac{d^2n_h}{dx^2} = \frac{q\mu_h}{\varepsilon_0\varepsilon_r}(n_h^2 - n_e n_h) - \mu_h \frac{dn_h}{dx}\frac{q}{\varepsilon_0\varepsilon_r}\left(\int (n_e - n_h)dx - E_{x0}\right) + L\frac{q(\mu_e + \mu_h)}{\varepsilon_0\varepsilon_r}n_e n_h - G$$

Eq 15

To solve equation 14 and 15 for calculating electron and hole densities inside three kinds of half devices, four boundary conditions regarding densities at edges (x=0 and 500 nm) are required. At the interface of the electron transport layer, electron density is considered to be zero based on experimental results of PL measurement (Most of carriers are quenched by charge selective layer). On the other hand, hole density is zero at the interface of the hole transport layer. Moreover, since there is no carrier diffusion at the boundary, density difference is regarded to be zero. Finally, we defined boundary conditions for the ETL/MAPbI$_3$, the HTL/MAPbI$_3$, and the glass/MAPbI$_3$ devices as shown in Supplementary Fig 2

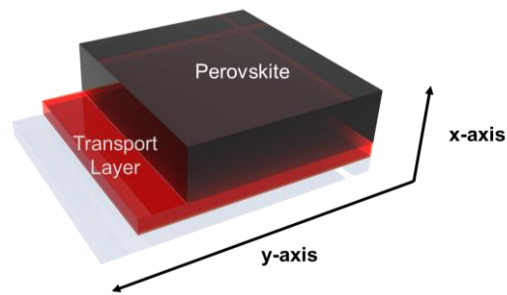

**Supplementary Information Figure 1.** Simplified 1-d model for Transport Layer/Perovskite using in this calculation

.

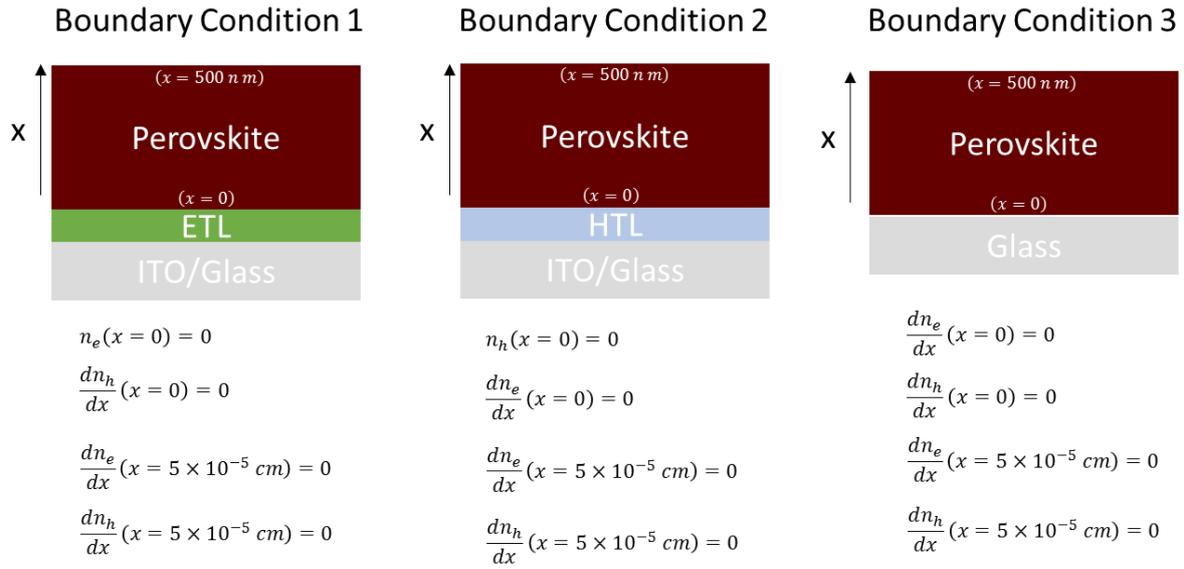

**Supplementary Information Figure 2**. Boundary conditions for ETL/MAPbI$_3$, the HTL/MAPbI$_3$, and the glass/MAPbI$_3$ devices to solve equation 14 and 15.

We could obtain numerical solutions for simultaneous equations 14 and 15 by using MATLAB.

Parameter values for the model are shown in Supplementary Table.1

| Parameter | Unit | Value |
|---|---|---|
| Q (unit charge) | C | 1.6 x 10$^{-19}$ |
| $\mu_e$ (electron mobility) | cm$^2$V$^{-1}$s$^{-1}$ | 5 |
| $\mu_h$ (hole mobility) | cm$^2$V$^{-1}$s$^{-1}$ | 3 |
| k (Boltzmann Constant) | JK$^{-1}$ | 1.38 x 10$^{-23}$ |
| T (Temperature) | K | 300 |
| L (Recombination constant) |  | 10$^{-5}$ |
| $\varepsilon_0$ (Vacuum Permittivity) | CV$^{-1}$cm$^{-1}$ | 8.854 x 10$^{-10}$ |
| $\varepsilon_r$ (Relative Permittivity) |  | 60 |
| G (Carrier Generation) |  | 1.7 x 10$^{21}$ |
| α (absorption coefficient) | cm$^{-1}$ | 1.48 x 10$^4$ |